# Analysis of multibunch free electron laser operation

Thorsten Hellert,[*] Winfried Decking, and Julien Branlard

*DESY, Notkestrasse 85, 22603 Hamburg, Germany*

(Received 8 May 2017; published 26 September 2017)

At the SASE-FEL user facilities FLASH and European XFEL, superconducting TESLA type cavities are used for acceleration of the driving electron bunches. The high achievable duty cycle allows for operating with long bunch trains, hence considerably increasing the efficiency of the machine. However, multibunch free electron lasers (FEL) operation requires longitudinal and transverse stability within the bunch train. The purpose of this work is to investigate the intra-bunch-train transverse dynamics at FLASH and European XFEL. Key relationships of superconducting rf cavity operation and the resulting impact on the intrabunch-train trajectory variation are described. The observed trajectory variation during multibunch user runs at FLASH is analyzed and related to both, intrabunch-train variations of the rf and the following impact on the multibunch FEL performance.

## I. INTRODUCTION

Single pass free electron lasers (FEL) are state-of-the-art technology to generate high brilliance self-amplified spontaneous emission (SASE) radiation [1], which is a powerful tool used for fundamental and applied research in many different fields of science. At the FEL user facilities FLASH (Free-Electron Laser in Hamburg) [2,3] and European XFEL (European X-Ray Free-Electron Laser) [4–6], the driving electron bunches are accelerated in superconducting radio-frequency (rf) resonators based on the TESLA (TeV-Energy Superconducting Linear Accelerator) [7] technology. Their main advantage over normal conducting cavities are the low Ohmic losses resulting in the possibility to combine high accelerating gradients with a high duty cycle, thus long rf pulse structure. The acceleration of long bunch trains allows for high bunch repetition rates and individual bunch patterns, adapted to the needs of the experiments. However, multibunch operation places additional demands on the machine, with respect to the beam stability within one bunch train.

Figure 1 illustrates the subject at FLASH. Plotted are the horizontal offset $\Delta x$ as recorded at two beam position monitors (BPM), upstream (left column) and downstream (center column) of the injector module as well as the SASE radiation energy $\Phi$ (right column) during a user run with bunch trains containing 400 bunches. The upper row shows the data of each bunch (black) within 5 min of machine operation and the mean value of each bunch train (red). In the mid row a time frame of 1 s is plotted, indicating the bunch trains with 10 Hz repetition rate. The bottom row exemplarily shows the data of one bunch train (black) and the corresponding 1 min mean of each bunch (red). A significant and systematic intrabunch-train trajectory variation occurs after the injector module. Its pattern is stable and its amplitude exceeds the bunch-to-bunch noise considerably. The same characteristics are noticeable for the SASE radiation energy. The SASE process critically depends on various beam parameters such as the peak current and transverse emittance of the bunches. However, in this paper we will show that the variation of the SASE radiation energy within one bunch train is caused primarily by intrabunch-train trajectory variations and will investigate their origin. Since the typical time scale within one bunch train is in the order of several 100 $\mu$s, fluctuations in iron magnets and microphonics can be excluded as their possible causes. Calculations on long-range wakefields [8], on the other hand, indicate that their amplitudes are too low to give a reasonable explanation for the observed trajectory variation. The dominating source of intrabunch-train trajectory variations are therefore expected to be variations of the rf parameters. The characteristics and constraints of superconducting cavity operation at FLASH and European XFEL are presented qualitatively in the following section. Thereafter a systematic study of multibunch operation at FLASH is presented, including rf-, trajectory-, and SASE radiation energy variations within one bunch train.

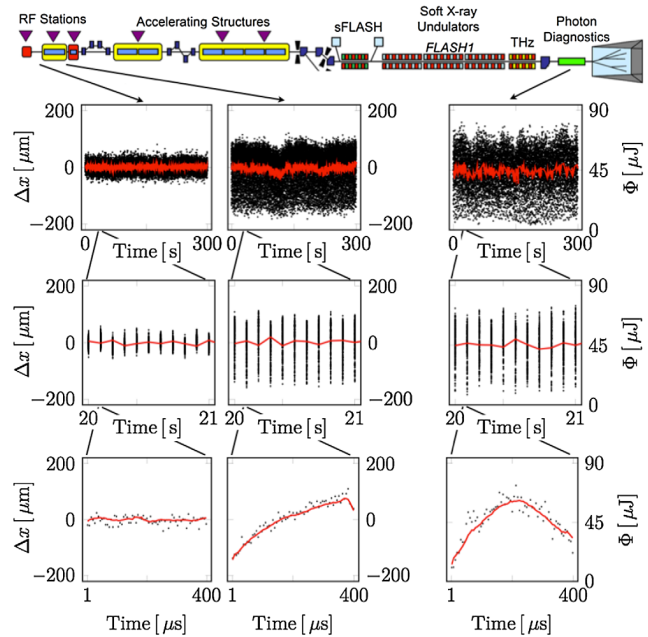

FIG. 1. Horizontal offset $\Delta x$ recorded at two BPMs, upstream (left column) and downstream (center column) of the injector module and the SASE radiation energy $\Phi$ (right column) at FLASH [9]. The upper and mid row show the data of each bunch (black) and the mean value of each bunch train (red) within different time frames of machine operation. The bottom row exemplarily shows the data of one bunch train (black) and the corresponding 1 min mean of each bunch (red).

---

[*]thorsten.hellert@desy.de

## II. PRINCIPLES OF SRF CAVITY OPERATION AT FLASH AND EUROPEAN XFEL

At FLASH and European XFEL, TESLA type cavities are used for accelerating the electron bunches. The TESLA cavity is a 9-cell standing wave structure of about 1 m length whose lowest TM mode resonates at 1.3 GHz. The cavity is built from solid niobium and is cooled by superfluid helium at 2 K. In contrast to the DC case, superconductors are not free from energy dissipation in rf fields. Since the heat loss exceeds the capability of a feasible cooling system in continuous wave (CW) operation at the required accelerating gradient at FLASH and European XFEL, the accelerator has to be operated in a pulsed mode. The rf is switched off periodically, thus long trains of bunches are accelerated in each rf cycle. A schematic drawing of the pulsed operation mode is shown in Fig. 2.

The maximum achievable accelerating gradient in a superconducting cavity is determined primarily by the quench limit, the quality factor $Q_0$ and the field emission of the cavity [10]. Due to disparate imperfections of the cavities, the operational limits of the accelerating gradient differ from cavity to cavity, for example in the range from 18 MV/m to 35 MV/m at FLASH [11]. Due to cost effectiveness several cavities are supplied by one high-power klystron. The output power is distributed in waveguides and then coupled into the cavities. A disadvantage of this setup is that the rf power control of each cavity cannot be done independently, but only within one rf station. A conservative approach in dealing with individual operational limits under these circumstances would be to set the accelerating gradient for all cavities within one rf station to the gradient limit of the poorest performing one. This would, however, reduce the efficiency substantially, given the above stated spread of operational gradient limits. There are different power distribution techniques available for optimizing the efficiency. The realization at FLASH and European XFEL fixes the waveguide distribution system through circulators, phase shifters, attenuators and (a-)symmetric shunt tees [11,12] so that the spread in power distribution matches the spread among cavity gradients within one rf station.

The primary role of the low-level rf (LLRF) control system is to stabilize the amplitude and phase of the accelerating field of the individual cavities within one rf station in order to assure a constant vector sum [13,14]. An algorithm calculates corrections to the driving signal of the klystron and applies piezo-based cavity tuning [15]. In addition to the feedback control loop which suppresses stochastic errors, adaptive feedforward correction is applied in order to compensate for repetitive perturbations [16]. As a result, the amplitude and phase stability of the vector sum of the accelerating field at FLASH are measured below 0.01% and 0.01° [17] within the bunch train, respectively. However, as will be described in the following, the accelerating field of individual cavities within one rf station shows significant intrabunch-train variations due to the effects of beam loading and Lorentz force detuning.

### A. rf slopes due to beam loading

A voltage counteracting the accelerating gradient is induced when a bunch of charged particles passes an accelerating cavity. This beam induced voltage drop is called beam loading [18] and is in good approximation proportional to the beam current. In order to assure a constant accelerating field for subsequent bunches, the beam loading has to be compensated by the rf power source. The loaded quality factor $Q_L$ of a cavity characterizes the amount of power coupled from the waveguide into the cavity. Choosing a $Q_L$-setting for a given distribution of operational limits and a vector sum based LLRF control system is a nontrivial matter. It is possible to find one optimal $Q_L$-value [19] for all cavities that provides the highest possible vector sum gradient. At the same time, it guarantees that the vector sum will remain flat under any beam or no beam condition, and that no cavity will quench during operation.

However, if the power is distributed differently between individual cavities, the accelerating gradient of individual cavities will remain flat only for one particular beam current. The effect will be exemplified for the given waveguide distribution of the sixth accelerating module at FLASH (ACC6) and the standard $Q_L = 3 \times 10^6$ for all cavities, using a cavity simulator [20]. The accelerating gradients of the cavities are calculated for different beam currents during the rfF flattop. Feedback is on, the vector sum is constant in all cases and there is no detuning considered. The results are shown in Fig. 3. The spread of gradients during the flattop apparent without beam (left) reflects the spread in operational gradients. For high beam current the individual cavity gradients show a significant slope during the flattop. The amplitude of these gradient slopes is proportional to the beam current and, moreover, proportional to the gradient spread, which is determined by the spread in operational gradients, thus the waveguide setup.

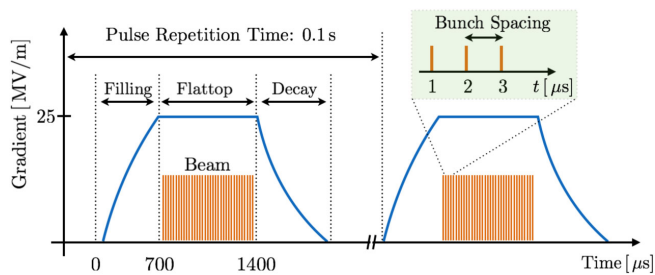

FIG. 2. Schematic drawing of pulsed rf operation with typical numbers at FLASH.

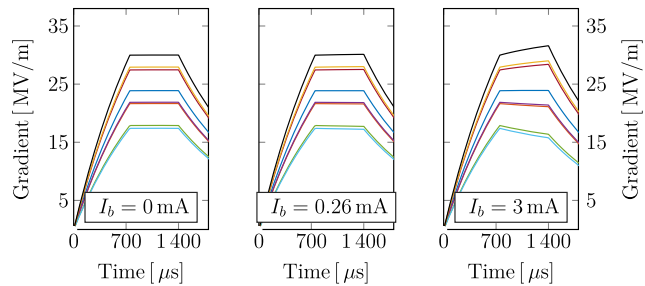

FIG. 3. Simulated accelerating gradients in ACC6 for a 650 $\mu$s flattop with the current FLASH $Q_L$-setup for different beam currents $I_b$. The vector sum is constant during the flattop. Feedback is switched on and there is no detuning considered. The different colored lines correspond to eight individual cavities.

The particular time dependence results from the fact that the beam loading is the same for all cavities, while the additional klystron power is distributed differently to the cavities. For an equal $Q_L$-setting the additional power is coupled by the same ratio into the cavities. The spread of gradients therefore increases during the beam duration.

The beam loading induced gradient slopes can be compensated for by adjusting the amount of power coupled into individual cavities. This was demonstrated for a two cavity case using motorized 3 dB hybrid [21]. Due to cost reasons, this level of control over individual cavity forward power is not available at FLASH or European XFEL. However, in the case of a fixed power distribution, the beam loading induced gradient slopes can still be compensated for by adjusting the $Q_L$-setup. It is worth noting that $Q_L$-correction is not necessarily possible for every beam current and gradient spread. In the scope of high beam current studies, $Q_L$-correction was successfully demonstrated at FLASH [22]. Implementation of an automated $Q_L$-correction, however, requires a careful balancing with other LLRF-feedback loops and until now there was no focus on this topic. As a result, the phase and amplitude of the accelerating field varies within one rf pulse, both at FLASH and European XFEL, depending on the beam current.

### B. rf slopes due to Lorentz force detuning

High electromagnetic fields in resonators lead to strong Lorentz forces on the walls of these structures. In order to ensure the cooling, the thickness and therefore rigidity of the walls cannot be chosen freely. As a consequence in a pulsed operation mode, the cavities are deformed dynamically in the range of some $\mu$m [13] at FLASH and European XFEL. This results in a dynamic behavior of the resonance frequency, a Lorentz force detuning (LFD). It scales quadratically with the accelerating field and due to the high $Q_0$, the detuning within one bunch train is in the order of the bandwidth of the cavity of about 300 Hz [23]. While coarse cavity tuning is performed using the cavity motorized step tuner, the LFD varies too fast to be compensated by conventional mechanical tuners [24]. At European XFEL fine tuning for each cavity will be handled by double piezoelectric elements installed inside a single mechanical support, providing actuator and sensor functionality or redundancy [25]. At FLASH, five out of seven accelerating modules are equipped with piezo-tuners.

As mentioned before, the rf klystron power is distributed to the individual cavities according to the waveguide setup. The Lorentz force detuning therefore differs for individual cavities. If the detuning changes, the amount of power coupled into or reflected from the cavity changes. The dynamic behavior of the coupling between the cavities and the power source can not be compensated for individual cavities by a vector sum rf control. As a result, the accelerating fields of individual cavities have a slope during the rf flattop. The effect will be shown qualitatively for ACC6 at FLASH using the cavity simulator. The accelerating gradients of the cavities are calculated for different coupling factors for the simulated Lorentz force, thus different amounts of cavity detuning within the flattop. In the simulation feedback is on, the vector sum is constant in all cases, $Q_L = 3 \times 10^6$, and there is no beam loading considered.

The results are shown in Fig. 4. The plotted values of the detuning $\Delta f$ are averaged over all cavities. For small detuning the effect of gradient slopes is negligible. For a mean cavity detuning of 200 Hz, on the other hand, a significant gradient slope is notable. The particular time dependence is due to the fact that the detuning scales quadratically with the accelerating field. Cavities with higher operational gradient limits will be detuned more than cavities with lower limits. This results in a further reduction of the coupling between the cavity and the rf power source. The additional power provided by the LLRF in order to assure a constant vector sum will be transferred less effectively into cavities with high accelerating gradient, hence higher detuning than into cavities with lower gradients. The spread of gradients therefore decreases during the flattop.

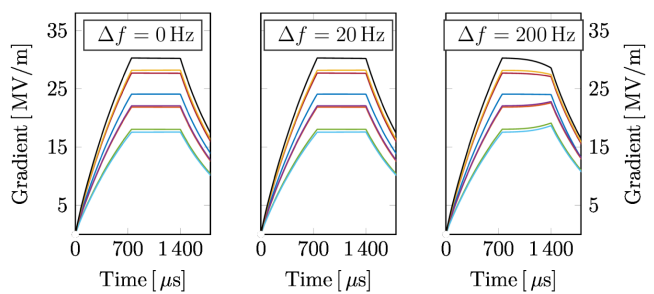

FIG. 4. Simulated accelerating gradient slopes in ACC6 for a 650 $\mu$s flattop for different mean cavity detuning $\Delta f$. The vector sum is constant during the flattop and there is no beam loading. The different colored lines correspond to eight individual cavities.

## III. MEASUREMENT RESULTS AT FLASH

This section provides a detailed evaluation of multi-bunch operation at FLASH. A total amount of 19 user runs with 400 bunches at the FLASH1 beam line between July 2015 and November 2016 have been systematically investigated. Although each machine setup is unique in its details, common features have been identified. Key aspects of the relation between intra-bunch-train trajectory variations and rf parameters are presented exemplarily on the basis of one set of data, taken at July, 29th 2015. Thereafter a general analysis of the multibunch FEL performance is given. All presented signals have been averaged over about 300 consecutive bunch trains with 10 Hz repetition rate. The beam current is 260 μA with a bunch spacing of 1 μs. Please note that all analyzed data is taken from user runs with stable machine operation, thus after deliberate SASE tuning.

### A. Intrabunch-train rf variation

The left plot in Fig. 5 shows exemplarily the measured detuning of individual cavities in the injector module at FLASH. The right plot shows the measured effect of Lorentz force detuning compensation achieved with piezo-tuners during the rf flattop at the third accelerating module [26]. Comparison with Fig. 4 reveals that the Lorentz force detuning compensation leads to a significant reduction of the slope of the accelerating gradient of individual cavities. Due to technical issues, however, the piezo-tuners at FLASH are by default switched off since 2015.

The resulting effect on the accelerating gradient is shown in Fig. 6. The left hand side shows the intrabunch-train variation of the amplitude of the accelerating field $\Delta V_0$ for individual cavities in the injector module. The mean amplitude is 19.5 MV/m. The amplitude variation of several 100 kV/m within on bunch train is typical at FLASH, as can be seen in the histogram on the right-hand side of Fig. 6. The rms value of the intrabunch-train amplitude variation is 245 kV/m. The mean amplitude of all 56 cavities is 17.5 MV/m for this special run, providing a final beam energy of about 960 MeV. In addition to the slow variation, small rf modulations with higher frequencies and a maximum peak-to-peak value of about 2 kV/m are noticeable. They are related to the excitation of $\frac{7}{9}\pi$- and $\frac{8}{9}\pi$-modes and treated in more detail in Ref. [27].

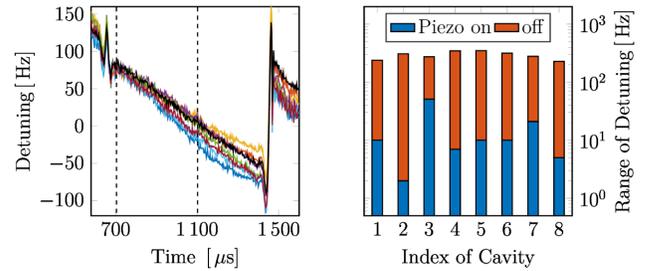

FIG. 5. Measured effect of Lorentz force detuning at FLASH. The left plot shows the detuning for individual cavities in the injector module without compensation. The rf flattop is from 600 μs to 1400 μs, the range between the dashed lines indicates the typical beam timing. The mean accelerating gradient is 16 MV/m. The right plot illustrates the effect of Lorentz force detuning compensation with piezo-tuners during the flattop at the third accelerating module at FLASH.

As described in the previous section, the slow variation of the accelerating field is considered to be dominated by the consequences resulting from Lorentz force detuning and beam loading. In order to classify their impact, simulations for the injector module using the cavity simulator are performed. The variation of the amplitude of the accelerating field within one bunch train is calculated and averaged over all cavities. If only Lorenz force detuning of maximal 100 Hz is considered, the mean variation is 65 kV/m. If only beam loading is considered, the mean variation is 16 kV/m. Considering both effects, the calculated mean variation is 52 kV/m. The lower value results from the fact that both effects counteract each other for one particular cavity (see Figs. 3 and 4). Both effects are appreciable in determining the slow variation of the accelerating field. However, Lorentz force detuning can be considered to be the dominating source at low beam currents which are typical for FLASH.

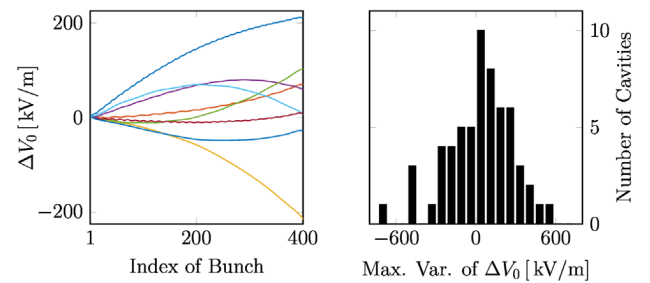

FIG. 6. Measured intrabunch-train variation of the amplitude $\Delta V_0$ of the accelerating field. The left plot shows the amplitude variation as a function of bunch index for eight cavities located in the injector module. The right plot shows a histogram of the maximum intrabunch-train amplitude variation of all 56 cavities at FLASH. The rms value is 245 kV/m.

It is worth noting that the particular amplitude and shape of the slow rf variation critically depends on the interaction of the LLRF parameters and the coarse tuning of all cavities within one rf station and can not be described generally at FLASH. In order to give a demonstration of this effect, rf data taken at five different user runs within one week of machine operation is compared. Beam current and the vector sum set point of the injector module differ by less than 1‰ within the data sets. However, the amount of intrabunch-train amplitude variation of the accelerating field of the first cavity of the injector module, for example, varies between 200 kV/m and 400 kV/m. The amplitude variation of the second cavity ranges from 70–210 kV/m. The variability is reproduced using the cavity simulator and changing the coarse tuning of individual cavities by several 10 Hz.

Small temperature drifts of the cavities are considered to dominate this variability, since the coarse tuning of individual cavities is not adjusted frequently during ordinary machine operation. However, the rms value of intrabunch-train amplitude variation of about 200–300 kV/m at FLASH as illustrated on the right-hand side in Fig. 6 is found throughout the data sets.

### B. Intrabunch-train trajectory variation

As discussed in detail in literature [28–30], accelerating cavities not only change the longitudinal momentum of the beam, but have an impact on its transverse motion, too. According to the Maxwell equations, transverse electromagnetic fields cannot be avoided off-axis. Their impact on the beam depends particularly on the misalignment of the beam trajectory in respect to the cavity axis and the rf parameters. If these parameters vary within one bunch train, the transverse dynamics of subsequent bunches will deviate from each other [27]. Since beam input trajectory and beam energy differ from cavity to cavity, the transverse forces induced by individual cavities will not cancel, even for a perfectly flat vector sum.

It is expected that the intrabunch-train trajectory variation is dominated by the previously described rf variations. The rf modulation pattern apparent in Fig. 6 will be mapped in some way on the trajectories, depending on the particular misalignments of the accelerating structures. In this paper we restrict the analysis on the primary effect. Higher order trajectory variations and such originating in the rf gun are discussed in more detail in Ref. [27].

In order to quantify the dominating intrabunch-train trajectory variation $\Delta\hat{u}$, frequencies below 10 kHz are isolated from each BPM reading $\Delta u$, where $u$ stands for $x$ and $y$, respectively. The range $\Delta\hat{u}_{\max} = \max(\Delta\hat{u}) - \min(\Delta\hat{u})$ is used as a measure of the amplitude of slow trajectory variation. The coefficient of determination $R^2 = 1 - \text{var}(\Delta u - \Delta\hat{u})/\text{var}(\Delta u)$ indicates the amount of trajectory variation which is expressed by the slow variation. For example, $R^2 = 0.9$ reveals that 90% of the apparent variance is covered by the low frequency signal.

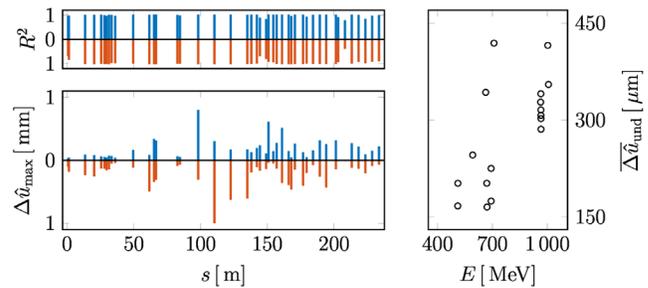

FIG. 7. Amplitude of slow intra-bunch-train trajectory variation, $\Delta\hat{u}_{\max}$ (lower left), and the corresponding coefficient of determination $R^2$ (upper left) for the horizontal (blue) and vertical (red) plane at each BPM. The right graph shows the mean offset variation $\overline{\Delta\hat{u}}_{\text{und}} = \sqrt{\Delta\hat{x}_{\text{und}}^2 + \Delta\hat{y}_{\text{und}}^2}$ at the entrance of the undulator section as a function of final beam energy $E$ for different user runs.

Figure 7 shows the results. The range $\Delta\hat{u}_{\max}$ for the horizontal and vertical plane and the corresponding $R^2$ value are plotted for each BPM. The large $R^2$ at most BPMs indicates that the low frequency signal covers most of the intra-bunch-train trajectory variation. The maximum amplitude of almost up to 1 mm is significant. A subsequent increase of the amplitude along the accelerating sections is noticeable. It is interpreted as a clear indication that the trajectory variations induced by the accelerating structures exceed the reducing effect of adiabatic damping. At a certain amount of misalignments and slope of the accelerating field, the induced trajectory variation of a cavity is larger than the reduction of the initial trajectory variation. Higher accelerating gradients, thus higher final beam energy, will therefore result in larger intrabunch-train trajectory variations. This consideration is supported by the results shown in the right graph of Fig. 7. The mean offset variation $\overline{\Delta\hat{u}}_{\text{und}} = \sqrt{\Delta\hat{x}_{\text{und}}^2 + \Delta\hat{y}_{\text{und}}^2}$ at the entrance of the undulator section is plotted as a function of final beam energy $E$ for different user runs. The intrabunch-train trajectory variation tends to increase with higher beam energy.

It is worth noting that this analysis remains partial, since only offsets are taken into account. A complete discussion should cover the whole phase space $[\Delta u, \Delta u']$. This is difficult since the trajectory tilt $\Delta u'$ is hardly measurable during ordinary user runs without significant errors [31]. However, when qualitatively comparing different user runs the trajectory offsets are a reasonable measure for trajectory variations.

### C. Intrabunch-train SASE performance

This paper focuses on the FLASH1 beam line, which is equipped with fixed (vertical) gap undulators. The properties of the SASE radiation are measured in a photon diagnostic section [32]. The total amount of SASE radiation energy $\Phi$, in the following referred to as SASE energy, is measured with a gas monitor detection system (GMD) between the undulator section and the experiments. At first a qualitative discussion on the multibunch SASE performance is made.

Figure 8 shows the relative variation of the SASE energy within one bunch train, $\Delta\Phi$, as recorded at the GMD during user runs with 400 bunches. The colored dots represent data available in the data acquisition system (DAQ) [33], the black crosses correspond to evaluated logbook [34] entries. In the left graph, $\Delta\Phi$ is plotted as a function of the beam energy $E$. The center graph relates $\Delta\Phi$ to the corresponding horizontal intrabunch-train offset variation at the entrance of the undulator $\Delta x_{und}$. In the right graph $\Delta\Phi$ is plotted as a function of the vertical offset variation $\Delta y_{und}$. Note that each point corresponds to a different user run.

The amount of intrabunch-train variation of the SASE energy is significant throughout the evaluated data. The lowest value, $\Delta\Phi = 14\%$, is reached for about $E = 500$ MeV, while the mean variation at beam energies around 690 MeV is $\Delta\Phi = 48\%$. At beam energies above 950 MeV the SASE energy varies more than 80% within one bunch train. From FEL theory it is expected that the sensitivity of the SASE process to small variations of beam parameters, for example the peak current or the initial beam trajectory offset, increases with higher beam energy. Additionally, the saturation length is reached later in the undulator. The observed increase of the variability of the intra-bunch-train SASE radiation energy with higher beam energy is therefore expected. However, Fig. 8 reveals that for the considered user runs the vertical intrabunch-train offset variation seems not significant in determining the multibunch SASE performance.

A quantitative discussion on the intrabunch-train variation of the SASE energy follows. Before its variability within the bunch train is related to trajectory variations, other effects are considered. The variation of the bunch charge within one bunch train is about 1% and is not a reasonable explanation of the observed radiation variability.

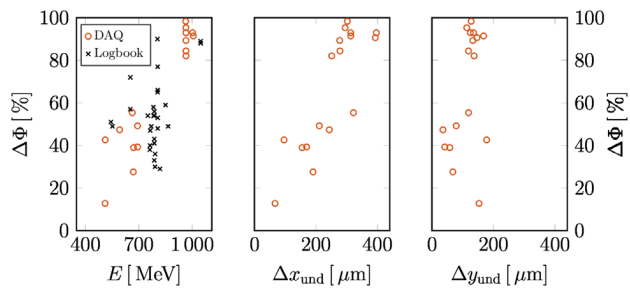

FIG. 8. Analysis of the intra-bunch-train variation of the SASE radiation energy, $\Delta\Phi$, recorded at FLASH during different user runs with 400 bunches. The right plot shows $\Delta\Phi$ as a function of the horizontal intrabunch-train offset variation $\Delta x_{und}$, measured at the entrance of the undulator section. In the mid graph $\Delta\Phi$ is plotted as a function of the vertical offset variation $\Delta y_{und}$. In the left graph, $\Delta\Phi$ is plotted as a function of beam energy $E$. The colored dots represent data available in the DAQ-system, the black crosses correspond to logbook entries.

However, the SASE process critically depends on the peak current of the electron bunch. The bunch compression monitors (BCM) [35] at FLASH suffer from lingering effects at 1 MHz bunch repetition rate. In order to resolve the bunch compression, thus an estimate of the peak current on a bunch-to-bunch scale, the signal of the BCM has to be corrected. The response of the involved electronics have been measured in 2017 and a correction algorithm was developed. However, it has to be noted that its validity for the considered user runs cannot be assured. Application of the lingering correction indicates that the intrabunch-train variation of the bunch compression is below 10% throughout the user runs. At FLASH the (slice-)emittance of single bunches within one bunch train can theoretically be measured using a transverse deflecting structure [36]. However, this is a comprehensive measurement that has not been made with 400 bunches. Regarding the analyzed user runs, the intrabunch-train variation of the emittance remains therefore unknown.

Based on the Ming-Xie model [37], a rough estimation is made. As an example the following values for the beam line at FLASH [38] are assumed for the calculation: beam energy 960 MeV, charge 260 pC, emittance 1.4 $\mu$m, and a peak current of 1.5 kA. The calculated radiation energy per bunch is 218 $\mu$J. Changing the peak current to 2.2 kA corresponds to a radiation energy of $\Phi = 263$ $\mu$J, thus an increase of 21%. Changing the emittance to 1 $\mu$m, on the other hand, leads to very similar values for the radiation energy. Increasing the peak current and decreasing the emittance simultaneously gives $\Phi = 296$ $\mu$J, which is an increase of 36%. All of the used values have been measured at FLASH during ordinary machine operation and reflect reasonable limits for the measurement.

However, the observed intrabunch-train variation of the SASE energy of more than 80% is unlikely to be explainable by a variation of the emittance and the peak current within the bunch train. In the unlikely case that both vary about 40%, the estimated variation of the radiation energy would be less than 40%. A more reasonable explanation for the observed intrabunch-train variation of the SASE energy is the horizontal intrabunch-train trajectory variation of several 100 $\mu$m, as will be demonstrated in the following.

Different user runs with similar beam parameters and comparable machine setups are identified. Six data sets with mean beam energy of about 965 MeV, which corresponds to a radiation wavelength of 6.7 nm, and four sets of data with a beam energy of about 680 MeV (13.5 nm) are analyzed. For each user run, the bunch with the highest SASE energy is found. Its radiation energy and trajectory are set to be the reference. The deviation of the SASE energy of other bunches is related to the deviation of their trajectory offset at the entrance of the undulator in respect to the reference bunch.

Time dependent GENESIS [39] simulations based on the recorded machine and beam parameters are performed.

Only trajectory variations are considered in each plane independently. Emittance and peak current, for example, are assumed to be constant throughout the simulations and there are no undulator errors and misalignments taken into account. The total amount of SASE energy at the exit of the undulator is calculated for different trajectory offsets at the entrance of the undulator section. The beam and undulator parameters for the simulation can be found in Table I.

Comparison between the experimental data and the simulations is shown in Fig. 9. The result confirms the previous considerations concluded from Fig. 8. The intrabunch-train variation of the SASE energy is clearly dominated by horizontal trajectory variations. At a beam energy of about 965 MeV, the average full-width-half-maximum value of the experimental data for the horizontal plane is 186 $\mu$m and 172 $\mu$m for the GENESIS results. At 680 MeV, the full width at 60% of the SASE energy is 317 $\mu$m and 330 $\mu$m for the data and simulation, respectively. The agreement is decent. For more detailed investigations on the multibunch SASE performance, important beam parameters and their variation within the bunch train remain unknown. However, it can be concluded that intrabunch-train trajectory variations can qualitatively and quantitatively explain the variation of SASE radiation energy within one bunch train. Based on the investigated experimental data at $E = 965$ MeV, the intrabunch-train offset variation must be limited to less than 60 $\mu$m to achieve an intrabunch-train SASE energy variation of less than 10%. At the concerned energy, the mean radiation energy of the bunch trains in the analyzed data is between 56% and 71% in respect to the reference bunch. Limiting the offset variation would therefore result in a significant increase of radiation energy of 20%–40% per bunch train. At 680 MeV, the mean radiation energy of the bunch trains is between 79% and 92%. Limiting the offset variation by a similar amount would result in an increase of radiation energy of 10%–20% per bunch train.

As discussed in this paper, intrabunch-train trajectory variations are a consequence of intrabunch-train variations of rf parameters. Regarding FLASH with low beam current and high accelerating gradient they are caused primarily by the effect of Lorentz force detuning. A limitation of the intrabunch-train rf variation by means of piezo-tuners should therefore improve the multi-bunch FEL performance considerably.

## IV. SUMMARY AND CONCLUSION

The SASE-FEL user facilities FLASH and European XFEL operate in a pulsed mode with long bunch trains, while each rf power station drives several cavities. A low-level-rf vector sum control is implemented, constraining the rf parameters of individual cavities. Individual operational gradients, Lorentz-force detuning and a fixed loaded quality factor setup result in significant intrabunch-train variations of rf parameters on the level of individual cavities: a slope of the accelerating field during the beam duration. In correlation with misalignments of accelerating structures, an intra-bunch-train trajectory variation is induced.

Analysis of the FEL performance of different user runs showed a significant intra-bunch-train variation of the SASE radiation energy throughout operations. Comparison with GENESIS simulations clearly suggested that intra-bunch-train trajectory variations are their dominant source. A significant improvement of the multibunch FEL performance at FLASH of about 30% at 7 nm radiation wavelength seems possible if intrabunch-train trajectory variations are limited sufficiently. Therefore, reducing the variation of rf parameters of individual cavities within one bunch train should be a considered objective for multibunch FEL operation.

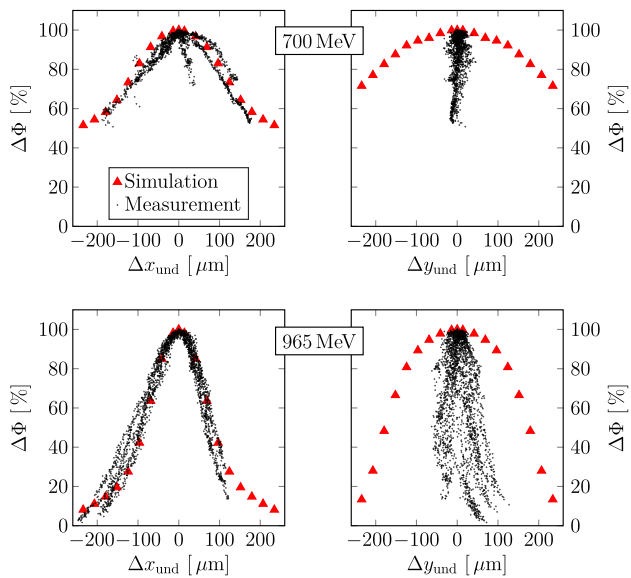

FIG. 9. Relative intrabunch-train variation of the SASE energy $\Delta\Phi$ as a function of the horizontal (left) and vertical (right) intrabunch-train offset, $\Delta x_{und}$ and $\Delta y_{und}$, respectively, at the entrance of the undulator at FLASH. The black dots represent data from several user runs with similar beam parameters at a mean beam energy of about 680 MeV (top) and 965 MeV (bottom). The red triangles are results of GENESIS simulations. For each user run the bunch with maximum SASE energy is used for reference by means of $\Delta\Phi$, $\Delta x_{und}$, and $\Delta y_{und}$.

## ACKNOWLEDGMENTS

This research was supported by the German Academic Scholarship Foundation. We would like to thank Bart Faatz, Sven Pfeiffer, Holger Schlarb, Sigfried Schreiber, Mathias Vogt, Klaus Flöttman, Florian Christie, and Jörg Rossbach for valuable comments that greatly improved the manuscript.

## APPENDIX SIMULATION PARAMETERS

See Table I.

TABLE I. Parameters for GENESIS simulations for the FLASH1 beam line.

| Electron beam | |
|---|---|
| Bunch charge | 260 pC |
| Transverse emittance | 1.4 $\mu$m rad |
| Peak current | 1.5 kA |
| Energy spread | 0.2 MeV |
| Undulator | |
| Type | planar |
| Gap | fixed |
| Period | 27.3 mm |
| Mean beta-function | 10 m |
| Undulator parameter $K$ | 1.23 |
| Segment length | 4.5 m |
| Number of segments | 6 |